\documentclass[aps,prl,twocolumn,showpacs]{revtex4-1}
\usepackage{amsmath}
\usepackage{graphicx}
\usepackage{units}  
\usepackage{xspace}
\usepackage{subfigure}
\usepackage{hyperref}
\newcommand{\beq}{\begin{equation}}
\newcommand{\eeq}{\end{equation}}
\newcommand{\bea}{\begin{eqnarray}}
\newcommand{\eea}{\end{eqnarray}}

\newcommand{\sx}{\sigma_{ x}}
\newcommand{\sy}{\sigma_{ y}}
\newcommand{\sz}{\sigma_{ z}}

\newcommand{\moire}{moir\'{e}\xspace}

\usepackage[usenames]{color}

\begin{document}
\bibliographystyle{apsrev}
 
\title{Landau Level Splitting in Rotationally Faulted Multilayer Graphene }

\date{\today}
\author{Hridis K. Pal$^1$, E. J. Mele$^2$, and M. Kindermann$^1$}
\affiliation{$^1$School of Physics, Georgia Institute of Technology, Atlanta, GA 30332-0430, USA\\
$^2$ Department of Physics and Astronomy, University of Pennsylvania, Philadelphia, PA 19104, USA}

\begin{abstract}
We show that valley degeneracy in rotationally faulted multilayer graphene may be broken in the presence of a magnetic field and interlayer commensurations. This happens due to a simultaneous breaking of both time-reversal and inversion symmetries leading to a splitting of Landau levels linear in the field. Our theoretical work is motivated by an experiment [Y.\ J.\ Song et al., Nature {\bf 467}, 185 (2010)] on epitaxially grown multilayer graphene where such  linear splitting of Landau levels was observed at moderate fields. We consider both bilayer and trilayer configurations and, although a linear splitting occurs in both cases, we show that the latter produces a splitting that is in quantitative agreement with the experiment.
\end{abstract}

\pacs{ 73.20.-r,73.21.Cd,73.22.Pr}
\maketitle 

The low-energy bandstructure of single-layer graphene is described by two Dirac cones located at two inequivalent points in the Brillouin zone (BZ), K and K$'$ \cite{neto:rmp09}. In the vicinity of these K and K$'$ points, which define the two bandstructure valleys of graphene, one finds quasiparticles with opposite chirality but degenerate in energy. This valley degeneracy is quite robust since it is guaranteed by two symmetries: time-reversal  and inversion.
Although the introduction of a magnetic field $B$ breaks the former, it does not break the latter, and  
the Landau levels (LL), therefore, are typically valley-degenerate. 
Experimentally it has been found that the valley degeneracy of LLs in graphene can be broken  in high magnetic fields \cite{zhang:prl06,jiang:prl07,checkelsky:prl08}. This phenomenon implies a spontaneous symmetry breaking, usually attributed to  electron-electron interactions \cite{goerbig:rmp11,nomura:prl06,yang:prb06,alicea:prb06,sheng:prl07,hou:prb10,kharitonov:prb12}.  

More recently, a splitting of LLs has been observed also in epitaxially grown graphene multilayers, likely, as well, due to a valley degeneracy breaking \cite{song:nat10}.  Such graphene multilayers typically behave  strikingly similarly to single layer graphene in many respects \cite{sadowski:prl06,orlita:prl08,miller:sci09}.  So one might surmise  that the observed lack of valley degeneracy is again  due to a spontaneous inversion symmetry breaking due to many-body interactions. In graphene multilayers, however, also the interlayer coupling may  lack  inversion symmetry. In this Letter we show  that consequently, and in contrast with single-layer graphene, the splittings observed in the experiment \cite{song:nat10}    do not require many-body interactions as their explanation, but  they can be understood quantitatively  as arising from the   interlayer motion of electrons.

The interlayer rotations in a graphene multilayer may be either commensurate or incommensurate. In the  incommensurate case one  has points where the hexagon centers of all layers are arbitrarily close to being coincident. Such points act as inversion symmetry centers to arbitrary precision. Any valley degeneracy breaking through interlayer coupling thus vanishes at large length scales. For commensurate interlayer rotations, however, such point symmetry centers do not necessarily exist. For bilayers with pairs of laterally coincident atoms it has been shown \cite{mele:prb10} 
that all commensurate rotations fall into one of two categories: sublattice exchange (SE) even, or SE odd. While SE even bilayers are inversion symmetric, SE odd bilayers are not.  In SE odd bilayers there is another symmetry that protects the valley degeneracy: mirror inversion at high symmetry lines with a subsequent interchange of the layers.
However, this symmetry may be broken by an interlayer bias $V$. In such a scenario the valley degeneracy of Landau Levels (LLs)
is then expected to be  lifted.

The experimental sample of Ref.\ \cite{song:nat10} consists of approximately six layers of graphene. The interlayer rotation angles are unknown except for the angle $\theta_{01}$ between the top two layers, which can be deduced from the \moire period observed in  topographic STM measurements: $\theta_{01} \approx 2.3 ^\circ$. For magnetic fields below  $B\simeq6\,{\rm T}$ the splitting in the first LL (LL$_1$) is found to be linear in $B$, and  of the order of one meV/T. Similar splittings are also observed in higher LLs. Although not quantitatively reported in the experiment, they appear to be smaller than the splitting of LL$_1$ (cf. Fig. 2 in Ref.~\cite{song:nat10}). In contrast, no splitting  of the zeroth LL    (LL$_0$) is observed. An attempt to attribute these splittings to the Zeeman effect fails since it would imply the same splitting for LL$_0$ as for all other LLs. Besides,  the observed magnitude of the splitting of LL$_1$ would require an electron $g$- factor of $g\approx 18$. Moreover, at high magnetic fields each split branch is observed to be further split into two more branches. This latter splitting is lesser in magnitude and yields  an electron $g$- factor of $g\approx  2$, which makes it reasonable to assume that the latter, in lieu of the former, is due to the electron spin. Most likely, therefore, the observed LL splitting at low fields, which is the subject of this Letter, is due to a valley degeneracy lifting. 
 
Also the mentioned spin splitting at high magnetic fields can  be enhanced  to a $g$-factor substantially larger than 2 (although still much smaller than 18) \cite{song:nat10}. This enhancement  occurs in a narrow energy range around the Fermi level and it is  likely due to electron-electron interactions. The low-field splitting of interest here, in contrast,  is observed   at energies comparatively far from the Fermi energy.  Many-body interactions are thus much less likely their origin. 
This motivates us to explore whether the observed low-field splitting can be explained within a non-interacting picture, as a result of valley degeneracy lifting due to broken inversion symmetry. 
To this end we consider theoretically SE odd bilayer and trilayer graphene configurations and predict a low-field splitting of LLs linear in the magnetic field, as observed in the experiment. While the splitting produced in the bilayer scenario is found to be much smaller in magnitude than that observed in the experiment,  we show that the LL splitting arising in both  studied trilayer configurations explains, within a non-interacting picture and for realistic parameters, all the salient features of the low field data in the experiment of Ref.\  \cite{song:nat10}.

\emph{SE odd commensurate bilayers}.---
At low energies the Hamiltonian of a graphene bilayer can be written as
$H_{\rm  bilayer}=H_0+H_1+(H^{\rm int}_{01}+h.c.)$,
where  $H_j$ is the Dirac Hamiltonian for layer $j$ and $H^{\rm int}_{jk}$ couples layers $j$ and $k$ (\emph{h.c} stands for hermitian conjugate). We write $H_j$ as $H_j= v(p_x\tau_z \sx+p_y \sy)+V_j$ with Pauli matrices $\sigma_\mu$ acting on the sublattice index and $\tau_\mu$ acting on the valley index. The potentials $V_j$ account for differences of the doping levels between the  layers 
and ${\bf p}$ is the canonical electron momentum. We choose for convenience $V_0=0$ and define $\sigma_+=\sigma_x+i\sigma_y$ and $\sigma_-=\sigma_x-i\sigma_y$. In the presence of a magnetic field $B$, 
$H_j=-\frac{i \omega_{\rm c}}{2}(a\sigma_{+}-a^{\dag}\sigma_{-})$
in valley K and $\sy H_j \sy$ in valley K$'$ (we set $\hbar=1$). Here  $\omega_{\rm c}=\sqrt{2} v/l_{B}$ is the cyclotron frequency, $l_{B} = 1/\sqrt{eB}$ is the magnetic length,   and $[a,a^\dag]=1$. 
The eigenstates of $H_j$  form Landau levels which are degenerate with wavefunctions $\psi_n = \left( \phi_{n-1},\phi_n\right)$ in valley K and $\psi'_n=\sy \psi_n$ in valley K$'$, 
$\phi_n$ being the LL wavefunctions of a conventional two-dimensional electron gas with quadratic dispersion and magnetic length $l_{B}$ \cite{neto:rmp09}. The energy of the $n$th Landau level  LL$_n$ is given by $\varepsilon_{n,\sigma}=\sqrt{n} \omega_{\rm c}$. 
 
 For SE odd commensurate bilayers the  interlayer coupling can be written as $H^{{\rm int}}_{01} = H^{\rm c}_{01}+H^{\rm nc}_{01}$,   with a momentum-conserving term \cite{mele:prb10,note1},
 \beq \label{HSO}
 H^{\rm c}_{jk}  = \frac{ {\cal V}_{jk}}{2} e^{i \vartheta_{jk}}\sigma_+ s_{jk},
    \eeq
and a term that does not conserve momentum \cite{lopes:prl07},
   \beq \label{Hnc}
 H^{\rm nc}_{jk} =  t_{jk}(\boldsymbol{r})s_{jk}.
 \eeq
Above, the matrices $s_{jk}$ transfer electrons from layers $k$ to $j$.  The nearest-layer coupling constant in Eq.~(\ref{HSO}) is approximately given by ${\cal V}_{jk} \approx \theta_{jk}^2 \gamma$ \cite{mele:prb10}, where $\theta_{jk}$ is the  rotation angle between layers $j$ and $k$  and $\gamma\approx 300\,{\rm meV}$ is the interlayer coupling for a Bernal stacked bilayer \cite{dresselhaus:aip02andothers}. At small  $\theta_{jk}\ll 1$ the interlayer matrix $t$ has long-wavelength components \cite{lopes:prl07}
\beq \label{t}
t_{jk}(\boldsymbol{r}) =\frac{\gamma}{3 }\sum_{p}e^{i \tau_z\delta\boldsymbol{K}^{(jk)}_{p}\cdot\boldsymbol{r}}\left[1+\frac{\zeta}{2}\left(e^{i \tau_z\phi_p}\sigma_++h.c.\right)\right],
\eeq 
with wavevectors $ \delta\!\boldsymbol{K}^{(jk)}_{p} =\boldsymbol{K}^{(j)}_{p}-\boldsymbol{K}^{(k)}_{p}$, where  $\boldsymbol{K}^{(j)}_{p}$ are the three equivalent $K$-points of valley $\nu=1$ in layer $j$ ($p=0,1,2$).  The parameter $\zeta$ parametrizes the sublattice asymmetry due to structural differences between different regions of the \moire pattern \cite{mele:prb11}
 ($\zeta=1$ in the absence of such asymmetry) and $\phi_p=2\pi p/3$.

Along the lines of Ref.\ \cite{kindermann:prb11} we next integrate out layer 1, obtaining an effective Hamiltonian $H_0^{\rm eff}(\omega)= H_0+ H^{\rm int}_{01}(\omega-H_1)^{-1}H^{{\rm int}\dag}_{01}$ for layer 0.  
In what follows we take the perturbative limit $\omega_c, \gamma \ll v \delta{K}$, 
which allows us to neglect  terms in $H_0^{\rm eff}$ that do not conserve momentum. The  momentum-conserving part of $H_0^{\rm eff}$ has two contributions:
$H_{0,{\rm c}}^{\rm eff}(\omega)=H_0+  \delta H^{\rm eff, nc}_{0,{\rm c}}+  \delta H^{\rm eff, c}_{0,{\rm c}}$,      
with   $ \delta H^{\rm eff, nc}_{0,{\rm c}}= H^{\rm nc}_{01} (\omega-H_1)^{-1}H^{{\rm nc}\dag}_{01}$ and $ \delta H^{\rm eff, c}_{0,{\rm c}} = H^{\rm c}_{01} (\omega-H_1)^{-1}H^{{\rm c}\dag}_{01}$. The contribution $ \delta H^{\rm eff, nc}_{0,{\rm c}}$ due to the interlayer coupling that does not conserve momentum    implies  a renormalization of $v$ \cite{lopes:prl07} besides other terms that are irrelevant in the low-energy limit $\omega_c \to 0$ (terms of higher   order in  ${\bf p}$). As expected,  $ \delta H^{\rm eff, nc}_{0,{\rm c}}$ does not break the valley symmetry. The part  $ \delta H^{\rm eff, c}_{0,{\rm c}}  $  due to $H_{01}^{\rm c} $ is, however, different: 
\beq \label{SOap}
\delta H^{\rm eff, c}_{0,{\rm c}}(\omega) =\frac{ (\omega-V_1) |{\cal V}_{01}|^2(1+\sz)}{ 2(\omega-V_1)^2-\left(2a^\dag a+1-\tau_z\right)\, \omega_{\rm c}^2 } .
  \eeq
 Both, the explicit valley-dependence in Eq.\ (\ref{SOap}) and the contribution $\propto \sz$, which breaks the A/B sublattice and thus inversion symmetry,    break the valley-degeneracy. 
In the limit 
$V_1 \gg  \omega_c \gg {\cal V}_{01}$, Eq.~(\ref{SOap}) may be simplified by expanding in $V_1^{-1}$, such that 
the    perturbative shift of LLs  to leading order in $\delta H^{\rm eff, c}_{0,{\rm c}}$ becomes $ \Delta \varepsilon_{n,\nu}= \langle \psi_{n,\nu}| H_{0,{\rm c}}^{\rm eff, c}(\varepsilon_{n,\nu})|\psi_{n,\nu}\rangle$, which splits the Landau levels by $ \Delta_n= \Delta \varepsilon_{n,1}- \Delta \varepsilon_{n,-1}$.
For $n=0$ we find  $\Delta \varepsilon_{0,\nu=1} ={\cal O}(V_1)^{-2}$ and $\Delta \varepsilon_{0,\nu=-1} =-|{\cal V}_{01}|^2/V_1+{\cal O}(V_1)^{-2}$ so that, to leading order,  LL$_0$ is split asymmetrically with
\beq
\label{SOsplit0}
 \Delta_0=|{\cal V}_{01}|^2/V_1+{\cal O}(V_1)^{-2}.
\eeq
For $n>0$ one has
\beq \label{SOsplitn}
  \Delta_n=2 \frac{|{\cal V}_{01}|^2v^2}{V_1^3} eB+{\cal O}(V_1)^{-4},
  \eeq
with the LLs split symmetrically, unlike  LL$_0$.

Equations (\ref{SOsplit0}) and (\ref{SOsplitn}) confirm that valley degeneracy in an SE odd bilayer with bias is indeed broken, resulting in a splitting of LLs. The splitting for all LL$_n$ with $n>0$ is linear in the field, consistent with the experiment. 
However, our model falls short in one major aspect:  the predicted splitting is significantly smaller than the experimental observation. With the  estimate ${\cal V}_{01} \approx \theta_{01}^2 \gamma$ \cite{mele:prb10}, the experimentally determined value $\theta_{01}\approx 2.3 ^\circ$, and  $V_1 \gtrsim \omega_c$, such that Eq.~(\ref{SOsplitn}) is valid, one  has $\Delta_n/B \lesssim 10^{-4} {\rm meV}/{\rm T}$, four orders of magnitude smaller than the one reported in Ref.\  \cite{song:nat10}. This quantitative disagreement originates mainly from the factor $\theta_{01}^2$  in ${\cal V}_{01}$ resulting in $\Delta_n\propto\theta_{01}^4$. 
Note that although the rotation angle between the top two layers is known to be small  in the experiment of  Ref.\   \cite{song:nat10}, the layers beneath the top two ones have unknown and possibly large rotation angles.  This motivates us to analyze the LL splitting  in graphene multilayers that results from  an SE odd commensuration with lower-lying layers.  As a minimal model we consider graphene trilayers.

\begin{figure}
\begin{center}
  \includegraphics[angle=0,width=0.9\columnwidth]{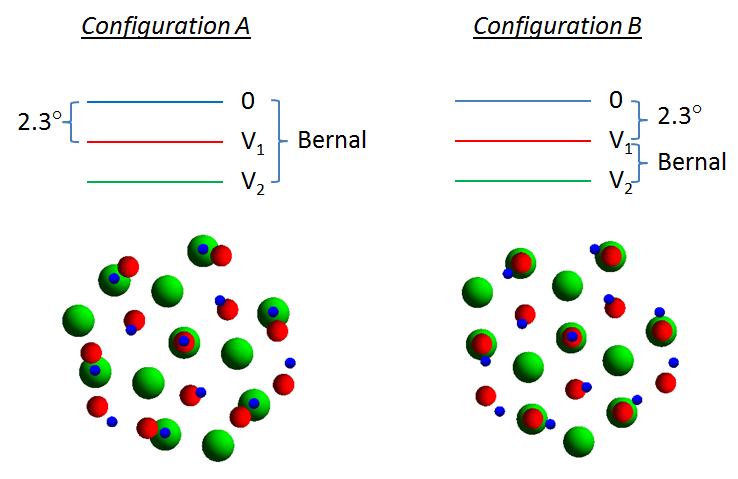}
  \caption{(Color online) The two trilayer configurations discussed in the text. Upper panel: schematic representation; lower panel: atomic positions for top (blue, small), middle (red, medium) and bottom (green, large) layers. The angle between top and middle layers are exaggerated for clarity.}
\label{fig0}
\end{center}
\end{figure}  

\emph{SE odd commensurate trilayers.}---In the absence of commensuration between the top two layers, there are two   configurations of graphene trilayers with broken inversion symmetry:
an SE odd-commensuration between either the top layer and the third layer counted from the top (configuration A) or the second layer and the third layer (configuration B)  (cf. Fig.~\ref{fig0}).  We assume Bernal stacking between each of these pairs since it produces the largest commensuration effects. Since Bernal stacked pairs of layers are known to occur with an appreciable probability of around $20 \%$ in epitaxially grown multilayers \cite{sprinkle:jpd10},  our model is not an unlikely scenario.

As in bilayers,  the low energy Hamiltonian of a graphene trilayer may be written as
$H_{\rm trilayer}=\sum_{j=0}^2 H_j+\sum_{j<k=0}^2 (H^{\rm int}_{jk}+h.c.).$
We begin with configuration A where the top layer and the third layer from the top are Bernal stacked with respect to each other. Then  $\theta_{01}=-\theta_{12}$ and we may neglect not only the momentum-conserving term $H^{\rm c}_{01}$, which has been found to be negligible in the previous section, but also  $H^{\rm c}_{12}$, which is of equal strength. We thus have $H^{\rm int}_{01}=H^{\rm nc}_{01}$, $H^{\rm int}_{12}={H}^{\rm nc}_{12}$, and  $H^{\rm int}_{02}=H^{\rm c}_{02}$, where  ${\cal V}_{02}=\gamma_1$ \cite{dresselhaus:aip02andothers},   the next-to-nearest layer coupling. Note that $H^{\rm int}_{02}$ conserves momentum under our assumption of Bernal stacking between layers 0 and 2, with $\theta_{02}=0$ (and an interlayer translation).  As before,  the terms in $H_{\rm trilayer}$ that do not conserve momentum do not lift the valley degeneracy. The analysis of the term of $H^{\rm eff}_0$ due to the momentum conserving  $H^{\rm int}_{02}$  is identical to that for bilayers. 
In the limit $\gamma_1 \ll \omega_c\ll V$ therefore, the LL splitting is still given by Eq.\ (\ref{SOsplitn}), but  with an enhanced coupling scale and a modified interlayer bias: ${\cal V}_{01}$ is replaced by ${\cal V}_{02}=\gamma_1$ and $V_1$ becomes $V_2$. The enhancement of the coupling constant leads to an increase in the magnitude of the splitting.  
\begin{figure}
\begin{center}
  \includegraphics[angle=0,width=0.8\columnwidth]{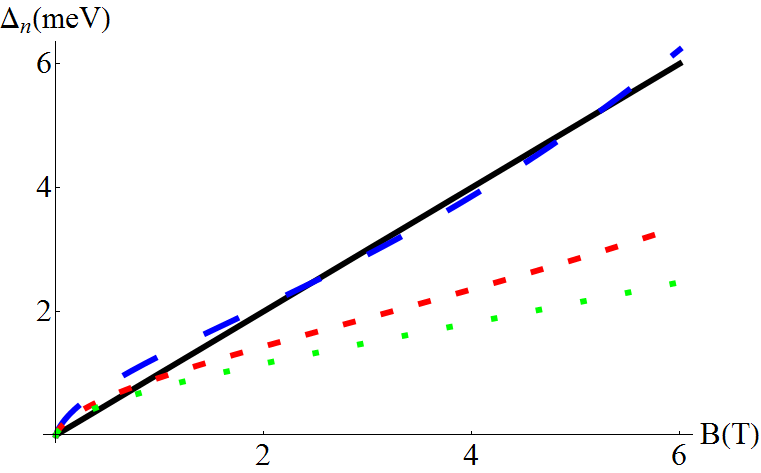}
  \caption{(Color online) Energy splitting  $\Delta_n$ of LL$_n$ for $n=1$ (blue, long-dashed), $n=2$ (red, medium-dashed), and $n=3$ (red, short-dashed) as a function of $B$ in trilayer configuration A for $\gamma_1 = \unit[15]{meV}$ and  $V= 40 \,{\rm meV}$. In black the experimentally determined $B$-dependence of LL$_1$: $\Delta_1=1 {\rm meV} B/{\rm T}$. }
\label{fig1}
\end{center}
\end{figure}

Returning to the experiment,  with the coupling $\gamma_1 \simeq \unit[15]{meV}$ found in the literature \cite{dresselhaus:aip02andothers}, a splitting of the observed magnitude, $\Delta_n/B \simeq 1 \mathrm{meV/T} $ is indeed obtained from Eq.\  (\ref{SOsplitn}) 
if we assume $V$ to be on the order of  tens of ${\rm meV}$,   a reasonable value. However, note that in this case   and for $B\gtrsim 1\,{\rm T}$ the condition $\gamma_1 \ll \omega_c\ll V$, required for the validity of  Eq.\    (\ref{SOsplitn}),  is not satisfied -- all three quantities are of the same order of magnitude.  Not surprisingly, the nonperturbative value of $\Delta_n$, calculated from the Hamiltonian Eq.\ (\ref{SOap}) along the lines of Ref.\  \cite{kindermann:prb11},  is generically not  linear in $B$ anymore.   Nonetheless, for some special choices of parameters in the above range of $\gamma_1$ and  $V$ the dependence is so close to linear  that it is consistent with the experimental observation. As an example we plot $\Delta_n$ in Fig.~\ref{fig1}  for   $\gamma_1 = \unit[15]{meV}$ and   $V= 40 \,{\rm meV}$. An almost linear magnetic field dependence of the energy of LL$_1$ with precisely the experimentally observed magnitude is predicted. Note that, unlike in the perturbative case, the exact calculation leads to a dependence of $\Delta_n$ on the LL index $n$: the splitting decreases with increasing $n$. Although not explicitly mentioned in Ref.~\cite{song:nat10}, this agrees qualitatively with the experimental observation as well(cf. Fig. 2 in Ref.~\cite{song:nat10}).

The splitting of LL$_0$ is found to be  $\Delta_0\gtrsim 20{\rm meV}$ for the above choice of parameters and $B>2 {\rm T}$. 
At first sight this appears to contradict the   experiment, which shows only one branch of LL$_0$. As discussed for bilayers, however, LL$_0$ is predicted to split asymmetrically, with one branch  unaffected and another shifted towards lower energy. 
In the experiment of Ref.\ \cite{song:nat10} LLs with negative indices were not observed. Their absence is attributed to them being far from the Fermi energy, incurring  extra inelastic broadening and attenuation. Similar reasoning can explain why the branch of LL$_0$, which is predicted to be shifted to low energies, is not observed in the experiment.  
Our results for this configuration, therefore, fit all the main features of the low field data of Ref.\ \cite{song:nat10}. 
We emphasize, however, that although the above parameters that yield an almost linear in $B$ dependence of the LL splitting  are realistic, the approximate linearity   requires fine-tuning in this scenario.

\begin{figure}
\begin{center}
  \includegraphics[angle=0,width=\columnwidth]{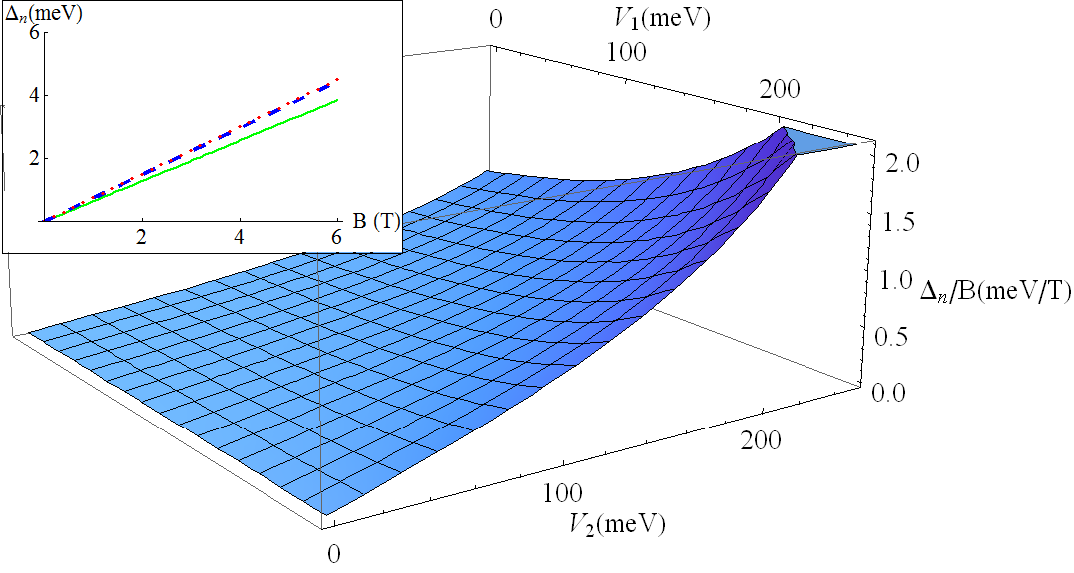}
  \caption{(Color online) $\Delta_n/B$ as  a function of $V_1$ and $V_2$. \emph{Inset}: $\Delta_n$ as a function of $B$ at $V_1=143 \mathrm{meV}$ and $V_2=170 \mathrm{meV}$ for calculations that are (i) perturbative  in${\cal V}_{12}$ with $\zeta=1$ [Eq.~(\ref{Dn}); solid (green)]; (ii)  non-perturbative  in ${\cal V}_{12}$ with $\zeta=1$ [dashed (blue)] ; (iii) non-perturbative  in ${\cal V}_{12}$ with $\zeta=1.2$ [dotted (red)]. }
\label{fig2}
\end{center}
\end{figure}

We finally turn to configuration B, where  the second and the third layer from the top are Bernal stacked with respect to each other. Now we have $\theta_{02}=\theta_{01}$, and following the same argument as in configuration A, we can neglect $H^{\rm c}_{01}$ and $H^{\rm c}_{02}$; thus,
$H^{\rm int}_{01}=H^{\rm nc}_{01}$ and $H^{\rm int}_{02}=H^{\rm nc}_{02}$.  The next-to-nearest layer  coupling  $\gamma_1 \simeq \unit[15]{meV}$  is much smaller than the   nearest layer coupling $\gamma\simeq 300 \,{\rm meV}$. This allows us to neglect also the effects of the  next-to-nearest layer    coupling Hamiltonian ${H}^{\rm nc}_{02}$, which are small quantitative corrections to those of $H^{\rm nc}_{01}$. The momentum-conserving   $H^{\rm int}_{12}=H^{\rm c}_{12}$ now has an inflated coupling scale ${\cal V}_{12}=\gamma$.  
The calculation for this configuration proceeds in two steps: we first integrate out layer 2. In our approximation, which neglects $H^{\rm int}_{02}$, this results in   a momentum-conserving effective Hamiltonian of layer 1: $ H^{\rm eff}_1 =H_1+ H^{\rm c}_{12} (\omega-H_2)^{-1}H^{{\rm c}\dag}_{12}$. Next we integrate out layer 1, arriving at an effective Hamiltonian for the top layer as before,  
$ H^{\rm eff}_{0} =H_0+ H^{\rm int}_{01} (\omega-H^{\rm eff}_1)^{-1}H^{{\rm int}\dag}_{01}$.
As for bilayers we first assume  $\omega_c, \gamma,V_j \ll v \delta{K}$ and  expand $H^{\mathrm{eff}}_0$ in $\gamma$ and $\omega_c$. The leading order effect is then again due to the momentum conserving part of $H^{\rm eff}_{0} $. Assuming $\zeta=1$ ($\zeta\neq 1$ discussed later) in $ H^{\rm int}_{01}$ [cf. Eq.~(\ref{HSO})] we find, 
 \beq \label{Dn}
\Delta_n=\frac{2 v^2 \gamma^4 V_2 (2v\delta K+V_1-V_2) (2v\delta K+V_1+V_2)}{3(v^2\delta K^2-V_1^2)^2(v^2\delta K^2-V_2^2)^2} eB,
 \eeq
for LL$_n$ with $n>0$. Unlike in configuration A, however, there is no splitting of LL$_0$: $\Delta_0=0$.

We now compare Eq.~(\ref{Dn}) with the experiment \cite{song:nat10}. With \cite{dresselhaus:aip02andothers} $\gamma = 300 \,{\rm eV}$,  $v\delta K= 450 \,{\rm meV}$ (corresponding to  $\theta_{01}=2.3^\circ$), and realistic values of  $V_1$ and $V_2$, we find $\Delta_n/B\simeq 1 \mathrm{meV/T}$, as in the experiment (cf. Fig.~\ref{fig2}). In this configuration, unlike in configuration A,  because of the clear scale separation $\omega_c\ll \gamma , v \delta{K}$, expansion in $\omega_c$ is justified and $\Delta_n$ is linear in the field  to a very good approximation without   any fine-tuning of parameters. Note that this means, unlike in configuration A, $\Delta_n$ is independent of $n$. A dependence on $n$ similar to that in configuration A can emerge, nevertheless, in  situations when $|V_{1,2}-v\delta K|\sim \omega_c$; however, such a calculation is outside the scope of this work. On the other hand, the assumption $\gamma \ll v \delta K$, underlying our derivation of Eq.~(\ref{Dn}), is  not strictly met in the experiment: $\gamma\lesssim v \delta{K}$. 
To address this partially we carried out a calculation that is nonperturbative in the coupling ${\cal V}_{12}=\gamma$  ($\gamma$ in $H^{\rm nc}_{01}$ is still accounted for perturbatively);  as shown in the inset of Fig.~\ref{fig2},   this  does not produce any qualitative effects, it merely increases the LL splitting  slightly. Higher order terms in perturbation theory in $H^{\rm nc}_{01}$ contain energy denominators $v \delta K_j \geq 2 v \delta K$ (up to corrections by $V_j$), so that  $\gamma/ v \delta K_j \lesssim1/3$ and our perturbation theory in $H^{\rm nc}_{01}$ appears to be justified, albeit marginally. Altogether this suggests that Eq.\ ({\ref{Dn}) does apply to a good approximation in the experimentally relevant parameter regime. 
Finally we investigate the effect of $\zeta\neq 1$. As seen in the inset of Fig.~\ref{fig2},  even a value of $\zeta=1.2$, which is on the large side of values found for honeycomb bilayer systems \cite{kindermann:prb12}, does not substantially change $\Delta_n$. While it does split  LL$_0$, this splitting   ($\le 3$ meV for $B\lesssim 6T$)  is of the same order of magnitude as the broadening of the state in the experiment of Ref.~\cite{song:nat10} (cf. Fig. 2 therein), making our prediction consistent with the apparent absence of a  splitting of LL$_0$ in that experiment.

\emph{Conclusion}.---We have shown that in the presence of a magnetic field, an interlayer bias, and interlayer commensuration of the SE odd type, valley degeneracy in multilayer graphene may be broken. We have demonstrated that for two generic trilayer configurations the resulting splitting of Landau levels can explain quantitatively such splitting observed at low fields in the experiment of Ref.\ \cite{song:nat10}.

\begin{acknowledgements}
We thank P.\ N.\ First and Y.\ J.\ Song for discussions and  acknowledge support by NSF under DMR-1055799 (HKP and MK) and   DOE under DE-FG02-ER45118 (EJM).
\end{acknowledgements}


\begin{thebibliography}{21}
\bibitem{note1} The form of Eq.\ (\ref{HSO}) follows from Ref.\ \cite{mele:prb10} through a rotation of the second layer by $60^\circ$, a symmetry of the bilayer. There is also the valley-reversed SE odd commensuration with interlayer coupling $\sx   H_{{\rm int}}^{\rm c} \sx$. It has the same valley splitting and we do not consider it separately.
\end{thebibliography}
  \end{document}